# Seamless Phase 2-3 Design: A Useful Strategy to Reduce the Sample Size for Dose Optimization

Liyun Jiang[1] and Ying Yuan[2*]
[1] Research Center of Biostatistics and Computational Pharmacy, China Pharmaceutical University, Nanjing, China
2 The University of Texas MD Anderson Cancer Center, Houston, TX, 77030
* Corresponding author: yyuan@manderson.org

## Abstract

**Purpose:**
The traditional more-is-better dose selection paradigm, developed based on cytotoxic chemotherapeutics, is often problematic When applied to the development of novel molecularly targeted agents (e.g., kinase inhibitors, monoclonal antibodies, and antibody–drug conjugates). The US Food and Drug Administration (FDA) initiated Project Optimus to reform the dose optimization and dose selection paradigm in oncology drug development and call for more attention to benefit-risk consideration.
**Methods:**
We systematically investigated the operating characteristics of the seamless phase 2-3 design as a strategy for dose optimization, where in stage 1 (corresponding to phase 2) patients are randomized to multiple doses, with or without a control; and in stage 2 (corresponding to phase 3) the efficacy of the selected optimal dose is evaluated with a randomized concurrent control or historical control. Depending on whether the concurrent control is included and the type of endpoints used in stages 1 and 2, we describe four types of seamless phase 2-3 dose-optimization designs, which are suitable for different clinical settings. The statistical and design considerations that pertain to dose optimization are discussed. The performance of these designs is evaluated by computer simulation.
**Results:**
Dose optimization phase 2-3 designs are able to control the familywise type I error rates and yield appropriate statistical power with substantially smaller sample size than the conventional approach. The sample size savings range from 16.6% to 27.3%, depending on the design and scenario, with a mean savings of 22.1%.
**Conclusion:**
The seamless phase 2-3 dose-optimization design provides an efficient way to reduce the sample size for dose-optimization trials and accelerate the development of targeted agents. Due to the interim dose selection, the phase 2-3 dose-optimization design is logistically and operationally more challenging, and should be carefully planned and implemented to ensure trial integrity.

**Keywords:** Dose optimization, Benefit-risk tradeoff, Seamless phase 2-3 design, Targeted therapy, Immunotherapy

## Introduction

Oncology drug development has been largely based on the more-is-better paradigm, where dose-finding studies are largely performed only in phase 1 clinical trials and intended to identify the maximum tolerated dose (MTD). This paradigm, developed based on cytotoxic chemotherapeutics, is often problematic for the development of novel molecularly targeted agents (e.g., kinase inhibitors, monoclonal antibodies, and antibody–drug conjugates). For targeted agents, increasing doses beyond a certain level may not enhance antitumor activity, and dose-limiting toxic effects may not be observed at a clinically active dose.[1] As noted by the Food and Drug Administration (FDA), the more-is-better paradigm "too often leads to selecting doses and schedules of molecularly targeted therapies that are inadequately characterized before initiating registration trials."[2]

In May 2021, based on the phase 2 portion of the CodeBreaK100 trial, the FDA approved sotorasib, the first drug targeting KRAS p.G12C mutation for metastatic non–small-cell lung cancers (NSCLCs).[3] However, the FDA required the sponsors to conduct a postmarketing trial to evaluate lower doses, since the clinical data show similar pharmacokinetic drug exposure, target saturation, and tumor response rates among patients treated with the dose used in the registration trial and those treated with lower doses.[1,4] This decision embodies a revolutionary paradigm shift in determining the labeled dose for targeted oncology drugs. To accelerate this paradigm shift, the FDA Oncology Center of Excellence initiated Project Optimus "to reform the dose optimization and dose selection paradigm in oncology drug development".[2] In May 2022, the FDA and the American Society of Clinical Oncology (ASCO) jointly sponsored a virtual workshop on "Getting the Dose Right: Optimizing Dose Selection Strategies in Oncology."

Shifting to a dose optimization paradigm requires investigators to evaluate and compare multiple doses in early phase trials to identify the optimal dose based on benefit-risk tradeoff before initiating registration trials. A straightforward approach is to take the dose-ranging-style method, where after the completion of the dose-escalation trial, two or more doses are selected based on exposure, target saturation, and other pharmacodynamic markers data collected early in clinical development[5]; they are then evaluated and compared in a randomized phase 2 trial in a fashion similar to dose-ranging trials that are routinely used in non-oncology drug development. This approach has been advocated by some researchers[6-10] and used in some trials[11-12].

One of the biggest challenges for this dose-ranging-style dose optimization is that it requires a larger sample size due to randomization and comparison among multiple doses. This is of particular concern for oncology as sponsors of oncology drugs frequently work under expedited timelines, and slow accrual is a common issue in oncology trials. One promising approach to address this issue is to use seamless phase 2-3 designs[13-20], which are named after their ability to combine an exploratory (stage 1) trial and a confirmative (stage 2) trial into one single trial. A fundamental feature of the phase 2-3 design is that at the end of the trial, stage 1 (i.e., phase 2) data will be combined with stage 2 (i.e., phase 3) data in an appropriate way to make inference on the treatment effect of the treatment or indication selected to continue to stage 2. In contrast, under the conventional trial design paradigm, phase 2 and phase 3 trials are conducted independently, and the evaluation of the treatment effect in phase 3 trials ignores the data from phase 2 trials. The phase 2-3 design is a type of novel adaptive design that has been embraced by regulatory agencies including the FDA and EMA. See the FDA's guidance on Adaptive Designs for Clinical Trials of Drugs and Biologics[21] for a trial example using a phase 2-3 design. Limited research has been done on using phase 2-3 designs for dose optimization.

The objective of this paper is to fill the gap and systematically investigate the operating characteristics of phase 2-3 designs as a strategy for dose optimization. We consider different phase 2-3 dose-optimization designs based on trial objectives and the type of endpoints (e.g., objective response rate (ORR) or survival endpoints such progression-free survival (PFS) or overall survival (OS)), and we particularly focus on design efficiency to reduce the sample size. We elucidate some unique characteristics and considerations pertaining to dose optimization and evaluate the performance of the designs by simulation.

## Methods

### Phase 2-3 Designs with Dose Optimization

For the purpose of dose optimization, the seamless phase 2-3 design consists of two stages (see Fig. 1). In stage 1, patients are randomized into $J$ ($\geq 2$) doses, with or without a control, to evaluate benefits and risks of the doses. $J$ doses are typically selected based on toxicity, pharmacokinetics/pharmacodynamics (PK/PD), and preliminary efficacy data collected in the phase I dose escalation study, which should demonstrate reasonable safety and anti-tumor activity. At the end of stage 1, an interim analysis is performed to select the optimal dose that produces the most favorable benefit-risk trade-off for further investigation in stage 2 of the trial. Stage 2 aims to confirm the efficacy of the selected optimal dose with a randomized concurrent control or historical control. Depending on whether the concurrent control is included and the type of endpoints used in stage 1 and 2, we distinguish four forms of phase 2-3 dose-optimization designs (see Figure 1) that are suitable for different clinical settings.

Design A includes a concurrent control in both stages, and it uses a short-term binary endpoint (e.g., ORR) in stage 1 to select the optimal dose and a long-term time-to-event endpoint (e.g., PFS or OS) in stage 2 to evaluate the therapeutic effect of the treatment. The use of a short-term endpoint in stage 1 facilitates a timely selection of the optimal dose to move forward to stage 2. A fundamental characteristic of Design A is that after evaluation of the short-term endpoint, patients in stage 1 will be continuously followed to evaluate their long-term time-to-event endpoint (e.g., PFS or OS); at the end of stage 2, data from both stage 1 and 2 will be used for the final analysis of PFS or OS. In contrast, the traditional paradigm conducts phases 2 and 3 independently, largely ignoring phase 2 data when analyzing phase 3 data. This is the key reason why a phase 2-3 design is more efficient and has the potential to reduce the sample size. Although not shown in the schema, when appropriate, stage 2 may include an additional interim futility/superiority analysis similar to the standard group sequential design[22-24]. Design B is a variation of Design A that includes only the control in stage 2 (see Design B), which could further reduce the sample size. The limitation of Design B is that, due to the lack of concurrent control in stage 1, it is challenging to combine stage 1 and 2 data to obtain an unbiased estimate of the treatment effect when there is a drift in the patient population. When a drift is unlikely, Design B is a reasonable choice. Design B was used in some clinical trials, e.g., a randomized multi-center trial of SM-88 in patients with metastatic pancreatic cancer[25], and RINGSIDE trial evaluating AL102 in desmoid tumors[26].

Design C is similar to Design A but simpler in that the same short-term endpoint (e.g., ORR) is used for (stage 1) interim analysis and (stage 2) final analysis. This design is appropriate when demonstrating an effect on the long-term endpoint (e.g., survival or morbidity) requires lengthy and sometimes large trials because of the duration of the disease course, and also the short-term endpoint is reasonably likely to predict clinical benefit on the long-term endpoint[27-29]. Design C was used in some clinical trials, e.g., a seamless phase 2-3 TNK-S2B trial of intravenous tenecteplase versus standard-dose intravenous alteplase for treatment of patients with acute ischemic stroke[30]. Design D is a simplified version of Design C without a control, which is appropriate in situations where there is a particularly acute unmet medical need (e.g., a refractory or resistant patient population) and/or the tumor under treatment is rare. Designs C and D are useful for drug development that targets accelerated approval from the FDA, which is often based on a short-term surrogate or intermediate clinical endpoint such as ORR. Similar to Design B, a limitation of Design D is the lack of concurrent controls, which may lead to a biased estimate of the treatment effect when there is a drift in the patient population.

In Designs A-D, there are two key decision points: one is at the end of stage 1 to select the optimal dose, and the other one is at the end of stage 2 to determine whether the treatment is effective compared to the concurrent or historical control. In what follows, we will describe design considerations pertaining to these two decision points.

**Selection of Optimal Dose**

At the end of stage 1, the optimal dose with the most desirable benefit-risk tradeoff is selected to move forward to stage 2 to further efficacy and safety evaluation. Depending on the trial setting, different metrics can be used to quantify the benefit-risk tradeoff and guide dose selection. For ease of exposition, we focus on a toxicity endpoint and an efficacy endpoint, noting that the selection of the optimal dose should also take into account all available data (e.g., pharmacokinetic/ pharmacodynamic (PD/PD) data). Given a dose $d$, let $p_E$ and $p_T$ denote its efficacy rate and toxicity rate, respectively. A straightforward measure for the benefit-risk tradeoff (or desirability) of $d$ is

$$U = p_E - wp_T,$$

where $w$ is a prespecified weight. This efficacy-toxicity probability-based desirability measure says that a one-unit (e.g., 10%) increase in the toxicity rate will incur a $w$-unit (e.g., $w$%) penalty in the efficacy rate. A larger value of $w$ favors selecting a dose with lower toxicity. When $w = 0$, $U$ leads to selecting the dose with the highest efficacy rate.

A more flexible approach is to use utility to quantify the benefit-risk tradeoff. This approach starts with specifying a utility score for each possible pair of (efficacy, toxicity) outcomes. For binary efficacy and toxicity endpoints, given a patient, there are four possible outcomes: 1 = (efficacy, no toxicity), 2 = (efficacy, toxicity), 3 = (no efficacy, no toxicity), 4 = (no efficacy, toxicity). Clearly, outcomes 1 and 4 are the most and least desirable cases; they are assigned the scores of $u_1$ = 0 and $u_4$ = 100, respectively. Using these two scores as references, the scores of outcomes 2 and 3 can be elicited from clinicians to reflect their clinical desirability (e.g., $u_2$ = 60 and $u_3$ = 40). When $u_2 > u_3$, it means that it is preferable to have efficacy at the cost of toxicity.

Given a dose *d,* let $\pi_1, \ldots, \pi_4$ denote the probability that a patient has one of the four outcomes. The desirability of *d* is the average of $u_1, \ldots, u_4$, weighted by $\pi_1, \ldots, \pi_4$ as follows:

$$U = \pi_1 u_1 + \pi_2 u_2 + \pi_3 u_3 + \pi_4 u_4.$$

The utility approach has several advantages. First, it aligns with clinical practice and thus is more accessible to non-statisticians. In our experience, it is often easier for clinicians to score the desirability of patient outcomes than quantify the tradeoff between efficacy and toxicity probabilities Second, the utility approach is more flexible and contains the efficacy-toxicity-probability-based approach as a special case when setting $u_2 + u_3$ = 100. Third, it is straightforward to handle endpoints with more than two levels (e.g., a three-level toxicity endpoint such as mild/moderate/severe) and more than two endpoints.

Given a measure of benefit-risk tradeoff, the optimal dose is selected as the dose that has the highest desirability to move forward to stage 2 for further evaluation of efficacy and safety.

To avoid the selection of toxic and/or futile dose, let $\phi_T$ and $\phi_E$ denote a prespecified toxicity upper limit (e.g., 0.33) and an efficacy lower limit (e.g., 0.2). We require that the optimal dose should satisfy the following safety and efficacy criteria:

$$\text{(Safety criterion)} \ \Pr(p_T < \phi_T | data) > C_T \quad (1)$$

$$\text{(Efficacy criterion)} \ \Pr(p_E > \phi_E | data) > C_E , \quad (2)$$

where $C_T$ and $C_E$ are thresholds that should be calibrated by simulation to ensure desirable operating characteristics. The safety criterion (1) says that based on the observe data, the probability that the toxicity rate is lower than $\phi_T$ should be great than the threshold $C_T$; and the efficacy criterion (2) says that based on the observe data, the probability that the efficacy rate is higher than $\phi_E$ should be great than the threshold $C_E$. In the case that none of the doses satisfy the safety and efficacy criteria, the trial should stop early after stage 1.

**Final Analysis of Significance**

The other key decision point is at the end of stage 2 to assess if the treatment effect of the selected optimal dose is significant compared to the concurrent or historical control. Under phase 2-3 designs, the simple approach of directly pooling stage 1 and 2 data and applying a standard testing method (e.g., log-rank or chi-square test) is problematic. This is because the dose evaluated in stage 2 is selected as the dose that has demonstrated the most promising treatment effect in stage 1, and thus stage 1-2 combined data are intrinsically biased toward a positive treatment effect, resulting in an inflated type I error. A number of statistical methods have been proposed to correct this type I error inflation, see Stallard and Todd (2011)[13] and Kunz et al. (2015)[14].

We here focus on a method known as the combination test with the closed testing procedure (CTCT)[15,16]. With CTCT, we apply a standard testing method (e.g., log-rank or chi-squared test) to stage 1 data and stage 2 data independently to obtain p-values, and then combined the p-values with appropriate weights and transformation. Based on the combined p-value, the closed testing procedure is used to determine if the result reaches statistical significance. The purpose of the closed test procedure is to adjust for the impact of interim optimal dose selection and control the familywise type I error rate (FWER), defined in the paragraph below, at the nominal level. The technical details of CTCT are provided in Supplementary Material A.

**Evaluation of Phase 2-3 Dose-optimization Designs**

The distinct structure of seamless phase 2-3 dose-optimization designs requires special consideration when evaluating their operating characteristics.  In the conventional paradigm, phase 2 and 3 trials are conducted separately, and thus we only need to control type I errors independently for each of them. Under the phase 2-3 dose-optimization design, however, the stage 1 dose selection is an integrated component of the design and influences the subsequent stage 2 testing. Each dose has a certain chance to be selected and tested against the control in stage 2, resulting in multiple testing issues and type I error inflation. One way to account for this is to control FEWR, defined as the probability of making one or more false discoveries (or type I errors) when performing multiple hypotheses testings.

The other unique feature of phase 2-3 dose-optimization designs is that the standard definition of power is not sufficient to characterize its operating characteristics, because the power does not consider the dose selection process. To this end, we introduce the *generalized* *power* to account for both (stage 1) selection and (stage 2) testing, defined as the probability of correctly selecting the optimal dose and also rejecting the null hypothesis (i.e., claiming that the treatment is effective) given that the treatment is truly effective. Besides the generalized power, it is also of interest to assess the performance of stage 1 dose selection. This can be quantified by the percentage of correction selection (PCS) of the optimal dose at the end of stage 1.

## Simulation studies

We evaluated the operating characteristics of Designs A-D using simulation. For each design, as shown in Figure 2, we considered three scenarios when two doses were evaluated in stage 1 (where scenario 0 represents the null case that all doses are ineffective, and scenarios 1 and 2 represent that dose 1 or 2 is optimal, respectively) and four scenarios when three doses were evaluated in stage 1 (where scenario 0 represents the null case that all doses are ineffective, and scenarios 1-3 represent that doses 1-3 are optimal, respectively). For stage 1, we considered that the null ORR = 0.2 versus the target ORR = 0.4, and the null ORR = 0.1 versus the target ORR = 0.3. For stage 2, the null and target hazard ratios were 1 and 0.64, respectively. For each design, sample sizes $n_1$ (stage 1) and $n_2$ (stage 2) per arm were chosen to achieve 80% generalized power. For example, given that the null and target ORR is 0.2 and 0.4, respectively, $(n_{1,}\, n_2) =$ (50, 100) for Design A, (50, 115) for Design B, (50, 80) for Design C, and (45, 30) for Design D when two doses are evaluated. When three doses are evaluated, $(n_{1,}\, n_2) =$ (80, 100) for Design A, (80, 110) for Design B, (90, 60) for Design C, and (80, 20) for Design D. In the simulation, we also studied how the population drift impacted the performance of Designs B and D, and evaluated how different choices of $n_1$ and $n_2$ impacted the design performance by varying the ratio of $n_1$ and $n_2$ with a fixed total sample size.

To demonstrate the potential gain of phase 2-3 designs in reducing the sample size, we compare each of Designs A-D with its conventional counterpart (denoted as CC) that consists of separated phase 2 and 3 trials that exactly match stage 1 and 2 of Designs A-D, respectively, but the determination of whether the treatment is effective is based only on phase 3 data. The sample size of the CC design was chosen to produce 80% generalized power. We conducted 10,000 simulations in each scenario and calculated the FWER, PCS, generalized power, and the average sample size of the trial. The detailed data generation methods and simulation settings were provided in Supplementary Materials B and C.

## Results

In null scenarios (i.e., scenario 0) with all doses being ineffective, Designs A-D control the FWERs under 5% (see Fig S2). Figures 3A and 3B show the PCS and generalized power. Across scenarios, Designs A-D yield PCS of above 80% (Fig 3A) and generalized power of around 80% (Fig 3B). Figure 3C shows the average sample size of designs A-D, which is substantially smaller than that of its CCs. The sample size savings range from 16.6% to 27.3%, depending on the design and scenario, with a mean savings of 22.1%. For example, for the two-dose setting, the sample size savings for scenarios 1 and 2 are 16.7% and 16.6% for Design A and 25.0% and 24.8% for Design D. It appears that the sample size savings are more for Design B than Design A, and more for Design C than Design D. For example, in two-dose scenario 1, the sample size savings of Design B are 4.8% higher than that of Design A, and in three-dose scenario 2, the sample size savings of Design C are 8.5% higher than that of Design D. The results are similar under different null and target ORRs (0.1 vs. 0.3); see Figure S4 and Supplementary Material D.

As expected, compared to Design A, Design B requires a substantially smaller sample size because of no concurrent control in stage 1. For example, in three-dose scenario 3, the sample size of Design B is 60.4 less than that of Design A. Similarly, Design D requires a smaller sample size than Design C. For example, in two-dose scenario 1, the sample size of Design D is 38.7% of that of Design C. The limitation of Designs B and D is that they lack a concurrent control at stage 1 or both stages, which may lead to a biased estimate of the treatment effect in the presence of patient population drift. Figure S5 in Supplementary Material E shows the FWER and generalized power of Designs B and D when there is a positive or negative drift. The positive drift leads to FWER inflation from 5% up to 13%, and a negative drift reduces the power from 80% to 70% in some cases.

We also studied how the sample size at stage 1 (i.e., $n_1$) affects the performance of the seamless designs, with a fixed total sample size. Figure 4 shows the simulation results for Design C under two-dose scenario 1 and three-dose scenario 2. The results for other designs and scenarios are similar (see Figure S6). The generalized power initially increases with $n_1$ and then starts to decrease after $n_1$ reaches a certain value. For example, in three-dose scenario 2, when $n_1$ increases, the generalized power increases up to around 80% when $n_1 = 80$, then achieves a plateau from $n_1 = 80$ to $n_1 = 110$, and thereafter starts to decrease. This shows that when using phase 2-3 trials, the allocation of sample size between two stages should be carefully considered and calibrated by simulation to achieve the highest efficiency.

## Discussion

We have systematically investigated the operating characteristics of seamless phase 2-3 designs for dose optimization under different trial objectives and endpoints. Our simulation studies show that compared to conventional approach, phase 2-3 designs can yield similar power and type I error control with a substantially smaller sample size, thus providing a useful approach to addressing the issues of limited accrual and tight timeline in oncology dose optimization trials.

Designs A-D can be readily extended in various ways to accommodate specific trial considerations. For example, in stages 1 and 2, interim futility analysis can be added to allow early stopping of a dose or the trial. Besides ORR, other novel continuous efficacy endpoints, such as ctDNA and minimal residual disease, can be used to better inform the selection of the optimal dose for stage 2. Also, at the end of stage 1, in addition to selecting the optimal dose, other adaption (e.g., sample size re-estimation) can be added when appropriate.

There are some logistical and operational challenges to using seamless phase 2-3 designs for dose-optimization. Due to design complexity, meetings about the trial design should be held between sponsors and regulatory agencies as early as possible in development to provide an opportunity for agencies to convey their expectations sooner, which potentially leads to more efficient studies. For example, the determination of the doses to be studied in phase 2-3 trials not only need the knowledge of therapeutic properties, patient populations heterogeneity, and whether any additional dose exploration is necessary for a supplemental application, but also the communication between the patients and providers[5]. Similar considerations also apply to the selection of the optimal dose when stage 1 completes.

To enable the selection of the optimal dose, seamless phase 2-3 trials require access to the unblinded data at the end of stage 1, which might compromise the trial integrity if not handled appropriately. The FDA guidance on adaptive designs[21] recommends that the access to comparative interim results should be limited to individuals with relevant expertise who are independent of the personnel involved in conducting or managing the trial, and who have a need to know. A Data Monitoring Committee (DMC) or a dedicated independent adaptation body is required to make the interim dose selection decision. Safeguards should be in place to ensure that the persons responsible for preparing and reporting interim analysis results to DMC are physically and logistically separated from the personnel tasked with managing and conducting the trial. This entails planned procedures to maintain and verify confidentiality, as well as documentation of monitoring and adherence to the operating procedures. To maximize the integrity of the trial, the investigators should prespecify the details of the design, including the anticipated number and timing of interim analyses, the criteria used for dose selection, methods used to control FWER and estimation of treatment effects, and the data access plan.

Other design strategies are available for dose optimization, e.g., phase 1-2 designs[31,32] that integrate phase 1 and 2 trials. Unlike conventional dose-finding designs that make dose escalation based on DLT and aim to find MTD, this class of designs performs dose escalation based on the benefit-risk tradeoff and aim to find the optimal dose, e.g., the EffTox design[33] or BOIN12 design[34]. As dose optimization and selection has been completed before onset of the confirmative trial, a phase 1-2 design approach simplifies the implementation of confirmation trials, which may be appealing in some cases.

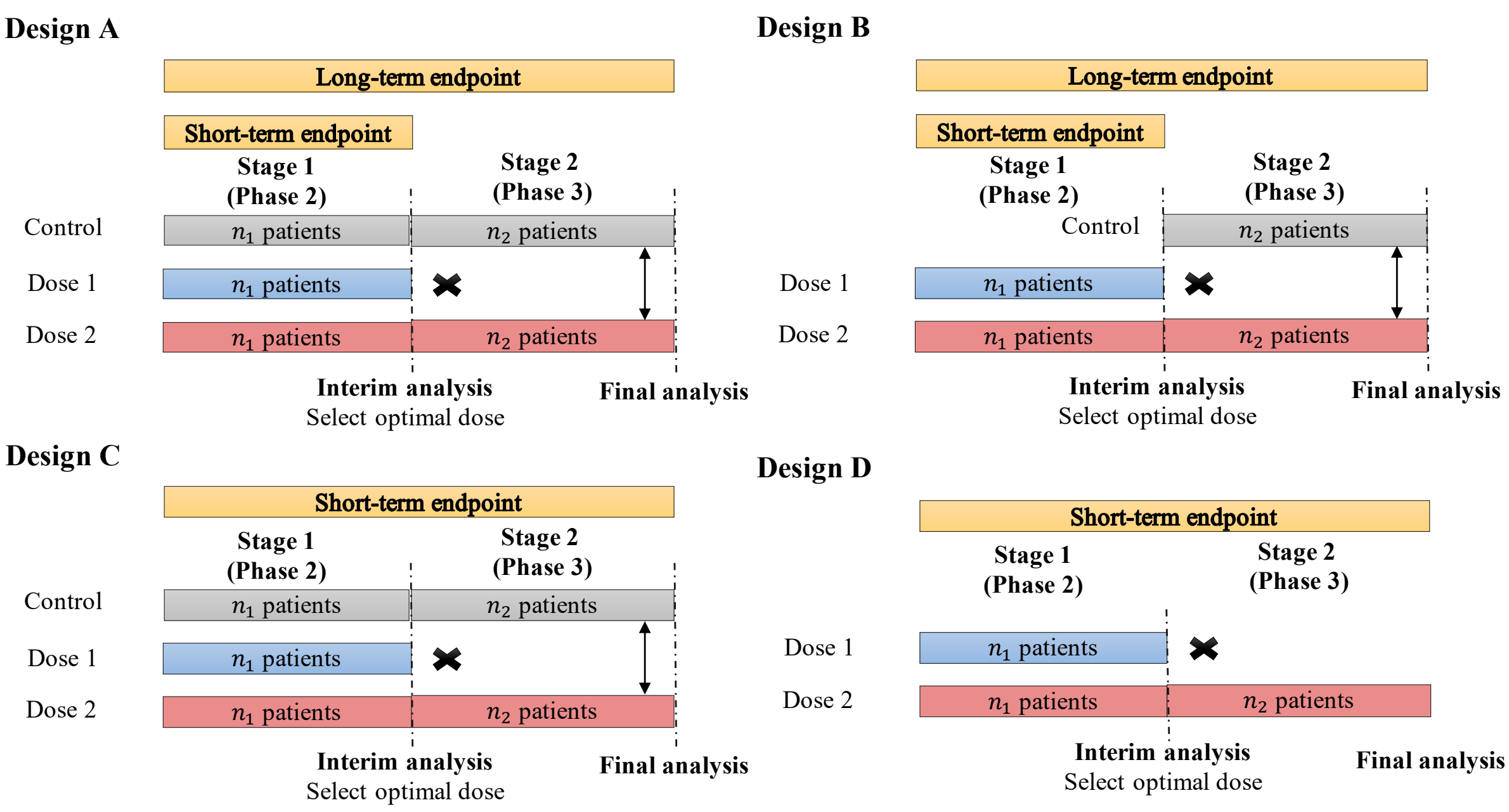

**Fig 1.** Schemas of the two-stage seamless phase 2-3 design. **Design A**: including a concurrent control in both stages, using a short-term binary endpoint (e.g., ORR) in stage 1 to select the optimal dose, and also using a long-term time-to-event endpoint (e.g., PFS or OS) in stage 2 to evaluate the therapeutic effect of the treatment. In this example, dose 2 is selected as the optimal dose at the end of stage 1, and patients are randomized to dose 2 and a concurrent control arm. **Design B**: a variation of Design A, including a control only in stage 2. **Design C**: similar to Design A but simpler in that the same short-term endpoint (e.g., ORR) is used for (stage 1) interim analysis and (stage 2) final analysis. **Design D**: a simplified version of Design C without a control in both stages.

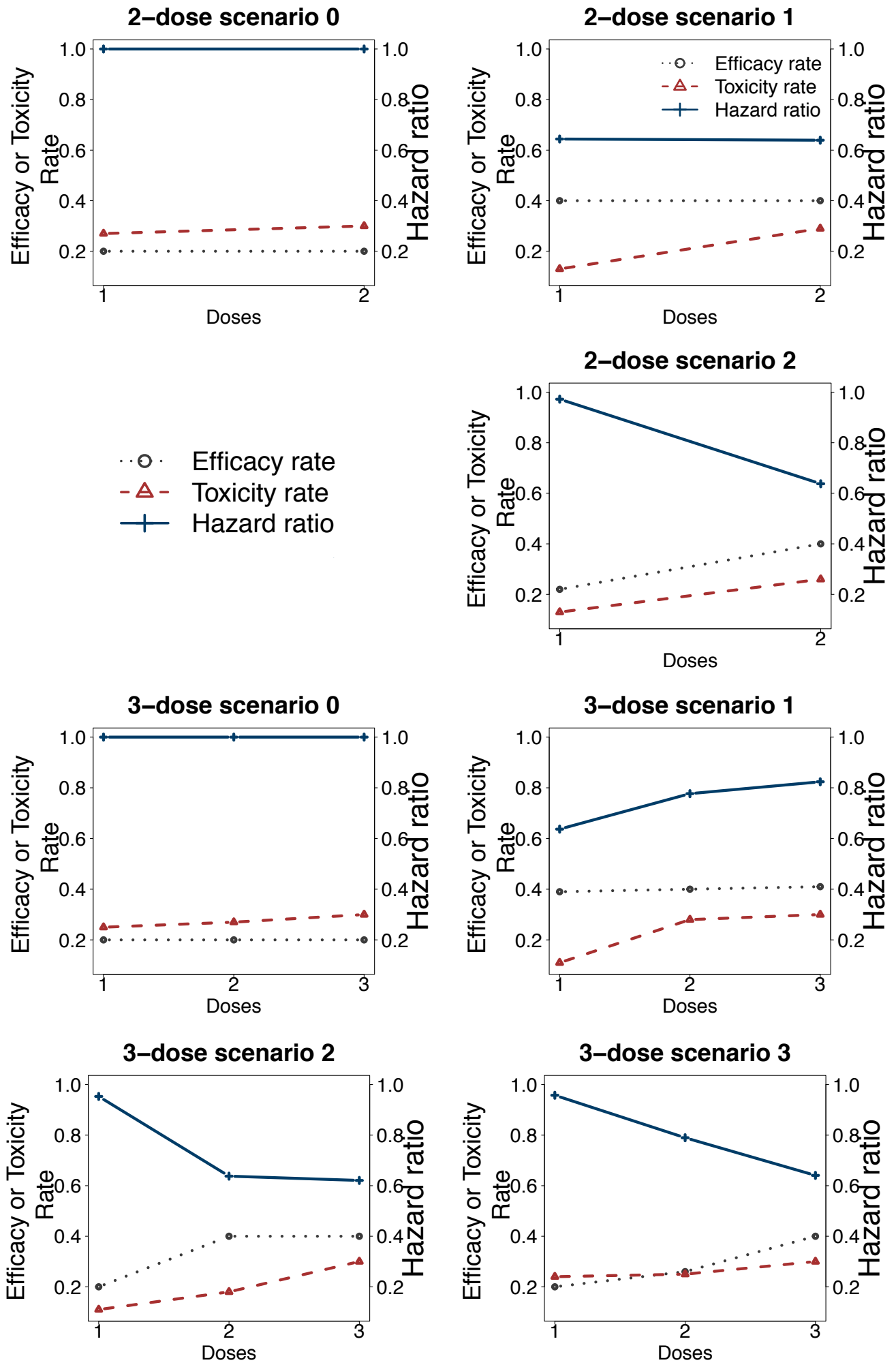

**Fig 2**. Response rate (grey dotted line), toxicity rate (blue dashed line) and hazard ratio (red solid line) for two-dose and three-dose scenarios. Scenario 0 represents no optimal dose, and scenarios 1-3 represent that the optimal doses are doses 1-3, respectively.

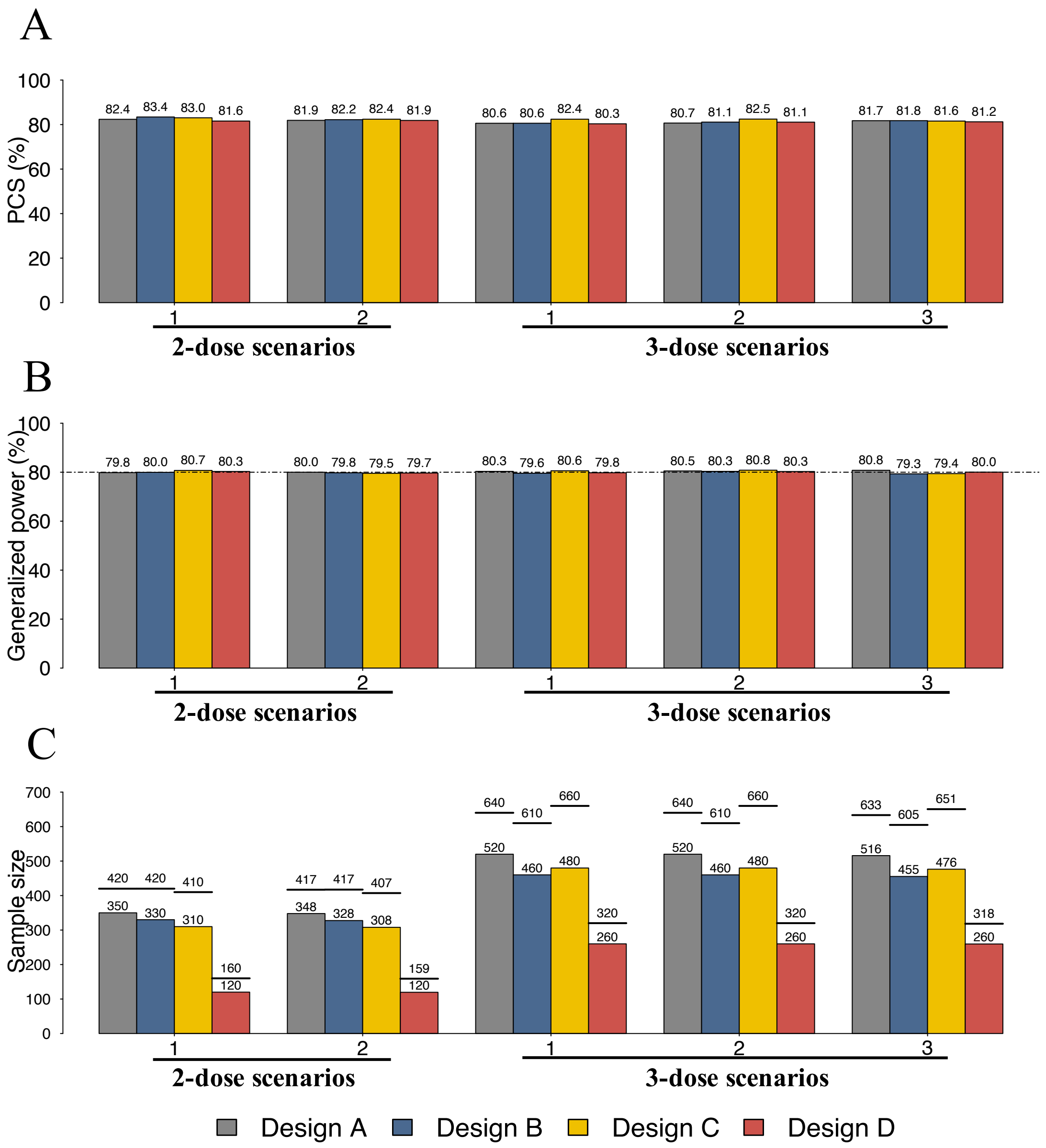

**Fig 3.** (A) The percentage of correct selection (PCS) of the optimal dose, (B) generalized power (%), and (C) the average sample size of Designs A-D, with the null and target ORRs of 0.2 and 0.4, respectively, and target hazard ratio of 0.64. In panel (C), the horizonal lines indicate the sample size of conventional approach.

A

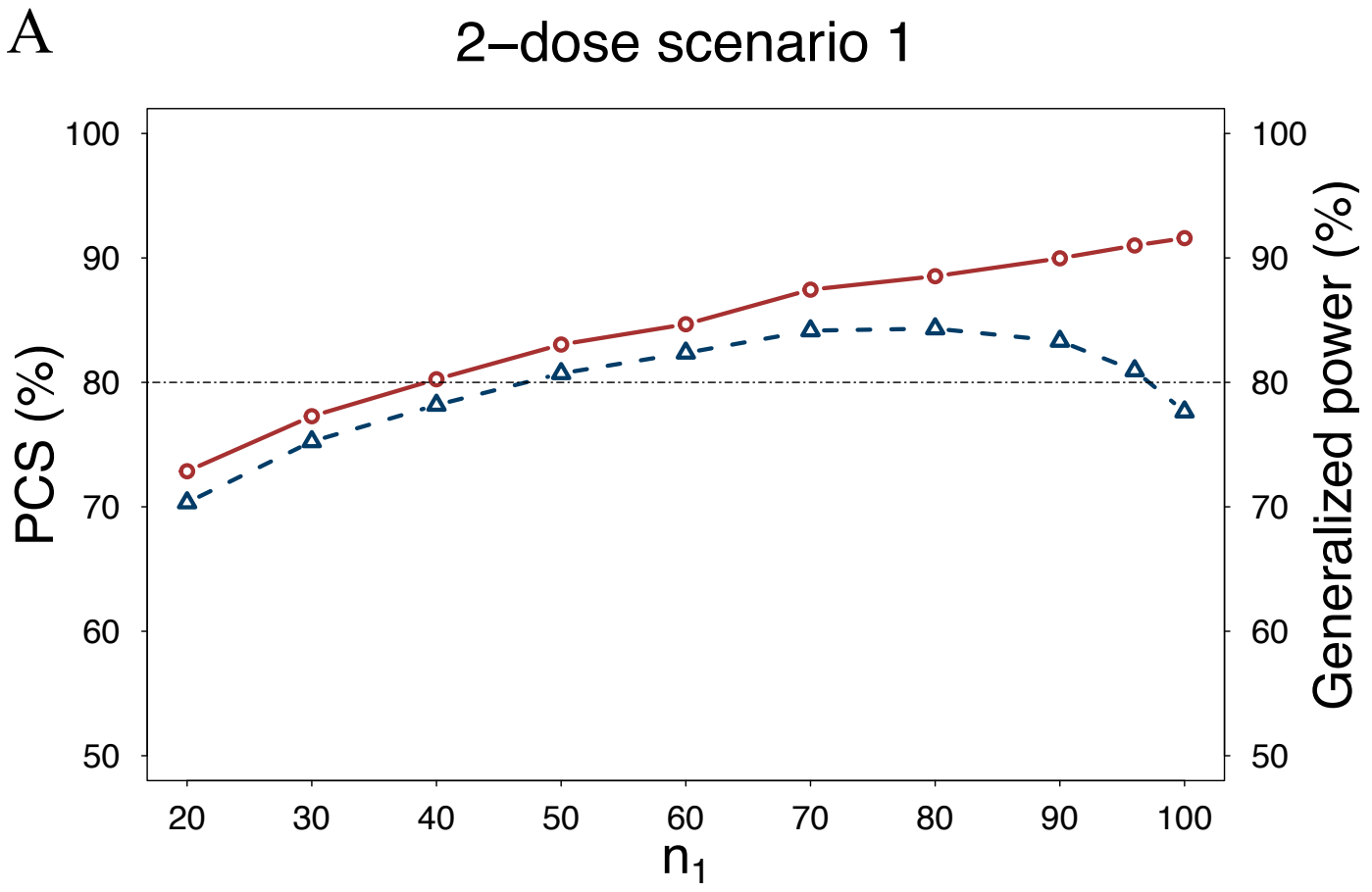

B

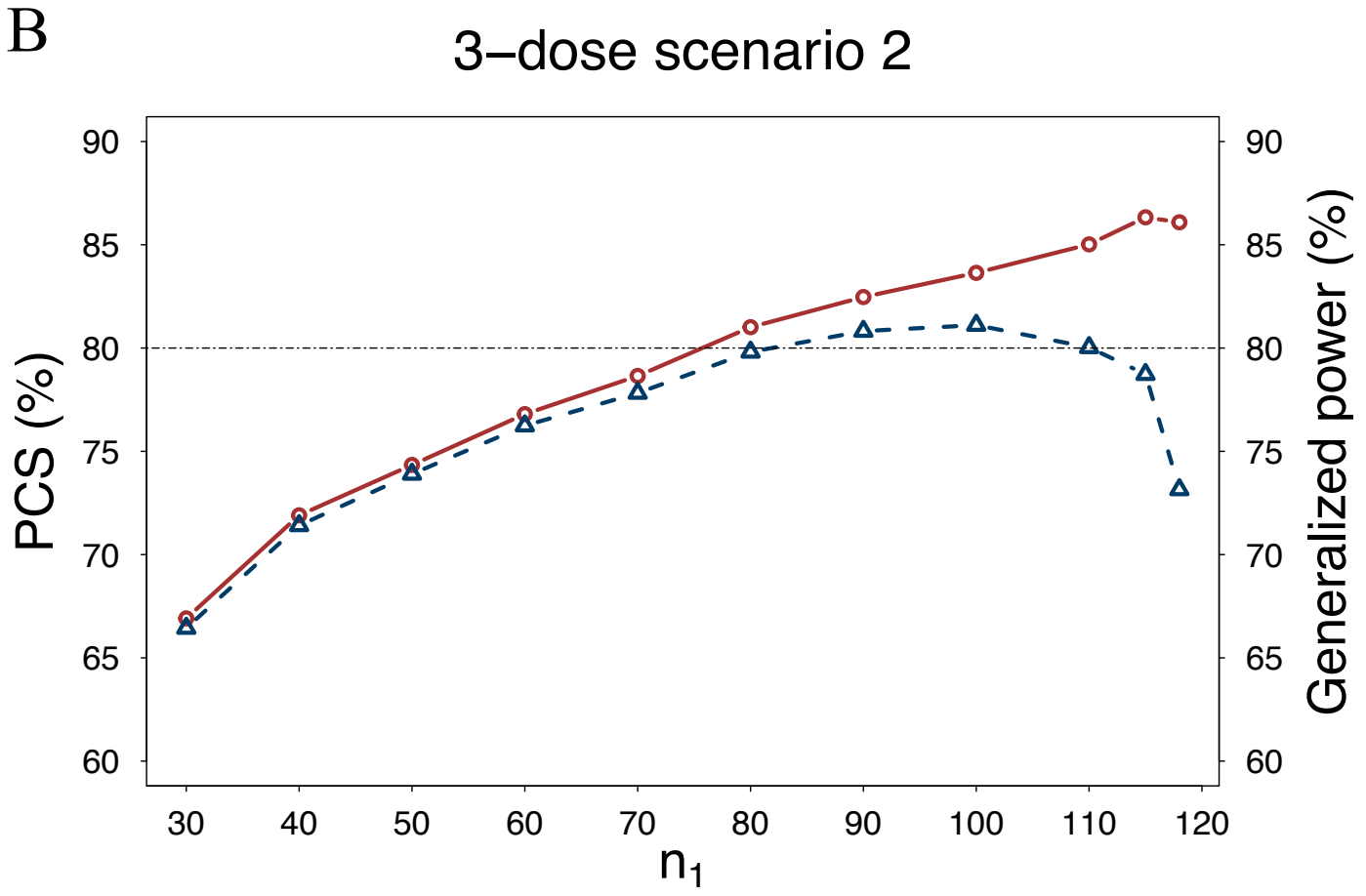

**Fig 4.** The percentage of correct selection (PCS) (red solid lines) of the optimal dose and generalized power (blue dashed lines) of Design C under different stage 1 sample size $n_1$ per arm in a two-dose scenario 1 and three-dose scenario 2. The total sample size is fixed.

# Supplementary Materials

## A. The combination test with the closed testing procedure (CTCT)

We assume the primary efficacy endpoint is a binary endpoint. For other endpoints, e.g., survival endpoints, the testing procedure is similar. Suppose $J = 3$ doses are selected to be evaluated. Let $p_{E,j}$ denote the efficacy rate of dose $d_j, j = 1, \cdots, 3$, and $p_c$ denote the efficacy rate of a concurrent control or a historical control $C$. The null hypotheses of interest are $H_j: p_{E,j} \leq p_c, j = 1, \cdots, 3$. Let $n_i$and $x_{ij}$ denote the sample size and the number of responders at $d_j$ in stage $i(= 1, 2)$, respectively. Suppose $P_{ij}$ is the p-value of comparing the dose $d_j$ with $C$ by using frequentist tests (e.g., t test) based on the efficacy data collected from stage $i$.

At the interim analysis (i.e., at the end of stage 1), we identify an optimal dose that yields the most favorable risk-benefit tradeoff by using the utility approaches. Note that the definition of interim decision rules (e.g., the dose selection rules) do not affect the familywise type I error rate (FWER), and there is no need to calibrate the interim decision rules to control FWER since the closed testing procedure will be used at final analysis to control FWER.

At the final analysis (i.e., at the end of stage 2), the combination test with the closed testing procedure is used. The closed testing procedure considers all seven hypotheses (including intersection hypothesis) $H_I = \bigcap_{l \in I} H_l$, where subset $I \subseteq \{1,2,3\}$. Any individual hypothesis $H_j$ is rejected if all $H_I$ with subset $I$ that includes $j$ are rejected at level $\alpha$. For example, $H_2$ is rejected if all $H_2, H_{12}, H_{23}, and\ H_{123}$ are rejected at level $\alpha$.

Without loss of generality, assume that at interim the dose $d_2$ is selected as the optimal dose to continue its evaluation in stage 2 of the trial. In this case, only those intersection hypotheses $H_I$ with $2 \in I$ remain to be tested, i.e., all hypotheses $H_2, H_{12}, H_{23}, and\ H_{123}$ need to be rejected at one-sided significance level $\alpha$. We need to calculate the test statistics $Z_I$, $I \in \{2, 12,\ 23, 123\}$ and obtain their critical values at level $\alpha$ to test the four hypotheses, which are described below. Let $P_{i,I}$ denote the p-value for a test of $H_I$ based on the data collected in stage $i\ (= 1, 2)$. At stage 1, $P_{1,I}, I \in \{2, 12,\ 23, 123\}$ are obtained from a Dunnett test[1]. At stage 2, since dose $d_2$ is selected as the optimal dose at interim, we compute $P_{2,2}$ based on data collected at stage 2, and define p-values for the intersection hypotheses by $P_{2,12} = P_{2,23} = P_{2,123} = P_{2,2}$. By using the combination test (e.g., the weighted inverse normal method[2]), the test statistics $Z_I$, $I \in \{2, 12, 23, 123\}$ to test the null hypothesis associated with $I$ at the end of the trial are

$$Z_I = \sqrt{\frac{n_1}{n_1 + n_2}}\Phi^{-1}(1 - P_{1,I}) + \sqrt{\frac{n_1}{n_1 + n_2}}\Phi^{-1}(1 - P_{2,I}),$$

where $\Phi^{-1}(\cdot)$ denotes the quantile of the standard normal distribution. The null hypothesis associated with $I$ will be rejected if $Z_I > \Phi^{-1}(1 - \alpha)$ (the $1 - \alpha$ quantile of the standard normal distribution). The null hypothesis of a lower or equal response rate for the optimal dose $d_2$versus $C$ will be rejected at multiple one-side $\alpha$ level, if all of intersection hypotheses $H_2, H_{12}, H_{23}, and\ H_{123}$ are rejected at the one-sided $\alpha$ level. The full procedure is shown in Figure S1.

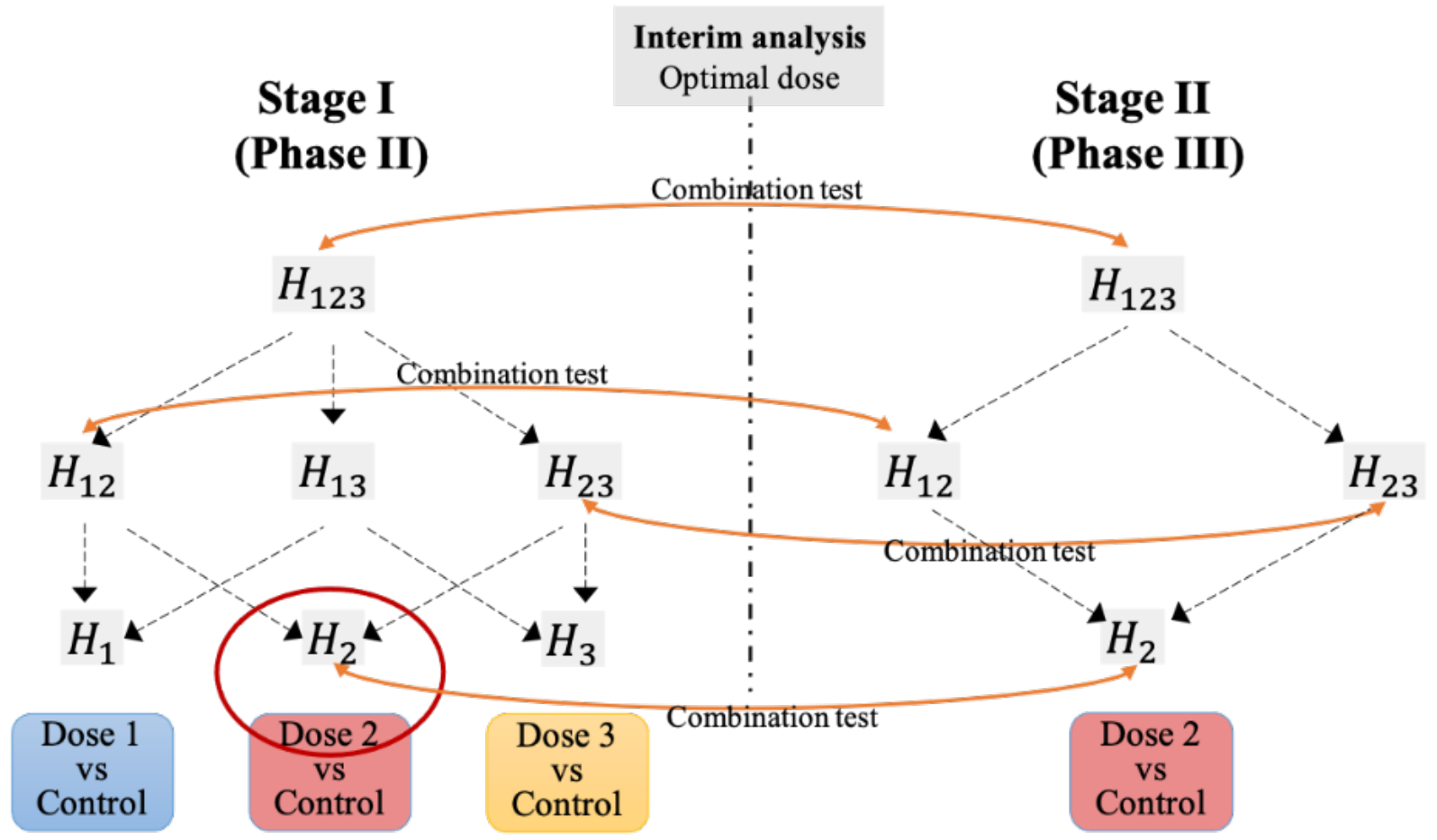

**Fig S1.** Illustration of closed testing procedure together with combination tests

## B. Data generation methods

Considering that the binary and survival efficacy endpoints are often strongly related, and that the binary endpoints can well predict the survival endpoints, we simulated the binary and survival data from a joint model. Specifically, we first simulated the binary outcome $Y_E$ for each patient from the Bernoulli distribution, and then, based on binary outcome $Y_E$, generated their survival times from the exponential distribution.

1) Control data

We first generated the control data. The binary outcome $Y_i$ of patient $i$ in the control arm was generated from $Y_i \sim Bernoulli(p_c)$, where $p_c$ is the response rate of the control arm. If $Y_i = 1\ (0)$, i.e., patient $i$ is in the responder (non-responder) group, we then generated the survival time $t_i$ of patient $i$ from an exponential model with hazard $h_1(t_i) = \lambda_1$ ($h_0(t_i) = \lambda_0$).

2) Data for dose $d$

We then generated the data for each patient at dose $d$. We first generated efficacy and toxicity data from a joint model. Specifically, for each patient at dose $d$, we first simulated a bivariate normal random variable $(z_T, z_E)$ with zero mean vector and the covariance matrix of $\begin{pmatrix} 1 & \rho \\ \rho & 1 \end{pmatrix}$, where $\rho$ measures the correlation between toxicity and efficacy. Let $\Phi^{-1}(\cdot)$ denote the inverse cumulative distribution of the standard normal variable, then the toxicity and efficacy outcomes for the patient are $Y_T = I\{z_T \leq \Phi^{-1}(p_T(d))\}$ and $Y_E = I\{z_E \leq \Phi^{-1}(p_E(d))\}$, respectively.

Given the binary outcome $Y_E$, we then simulated the survival time $t$ from a proportional hazard model. If $Y_E = 1$ (i.e., patient is in the responder group), we simulated $t$ from

$$h(t) = h_1(t)exp\{\beta_1\}, \ \ (responders)$$

and if $Y_E = 0$ (i.e., patient is in the non-responder group), we simulated $t$ from

$$h(t) = h_0(t)exp\{\beta_0\}, \ \ (non-responders)$$

where $h_0(\cdot)$ and $h_1(\cdot)$ are the hazards of the non-responders and responders, respectively, in the control arm, which were described previously. $exp\{\beta_1\}$ ($exp\{\beta_0\}$) was the hazard ratio between the responder (non-responder) group in the $d$ arm and responder (non-responder) group in the control arm.

**C. Simulation settings**

We considered two or three doses and compared them with a concurrent control arm or a historical control. We considered that the null ORR = 0.2 versus the target ORR = 0.4, and the null and target hazard ratios were 1 and 0.64, respectively. The accrual rate was two patients/month. The toxicity upper limit $\phi_T = 0.3$ and the efficacy lower limit $\phi_E = 0.3$. We considered the correlation between toxicity and efficacy $\rho = 0$. The thresholds $C_T = 0.1$ and $C_E = 0.1$. The scores in utility were set by $u_1 = 0, u_2 = 40, u_3 = 60, u_4 = 100$, which is identical to the efficacy-toxicity-probability-based approach, i.e., $U = p_E - wp_T$ with $w = \frac{u_2}{u_3}$. For each design, sample sizes $n_1$ and $n_2$ per arm were chosen to achieve 80% generalized power. For Design A, $n_1 = 50, n_2 = 100$ (for two doses) and $n_1 = 80, n_2 = 100$ (for three doses); for Design B, $n_1 = 50, n_2 = 115$ (for two doses) and $n_1 = 80, n_2 = 110$ (for three doses); for Design C, $n_1 = 50, n_2 = 80$ (for two doses), and $n_1 = 90, n_2 = 60$ (for three doses); and for Design D, $n_1 = 45, n_2 = 30$ (for two doses), and $n_1 = 80, n_2 = 20$ (for three doses). We considered three scenarios with two doses evaluated in stage 1, where scenario 0 represents the null case (all doses are ineffective), and scenarios 1-2 represent that dose 1-2 are optimal, respectively; and four scenarios with three doses evaluated in stage 1, where scenario 0 represents the null case, and scenarios 1-3 represent that doses 1-3 are optimal, respectively. For each dose, the true efficacy rate, toxicity rate, and hazard ratio are shown in Figure 2. FWERs are shown in Figure S2. PCS, generalized power, and sample size are shown in Figure 3.

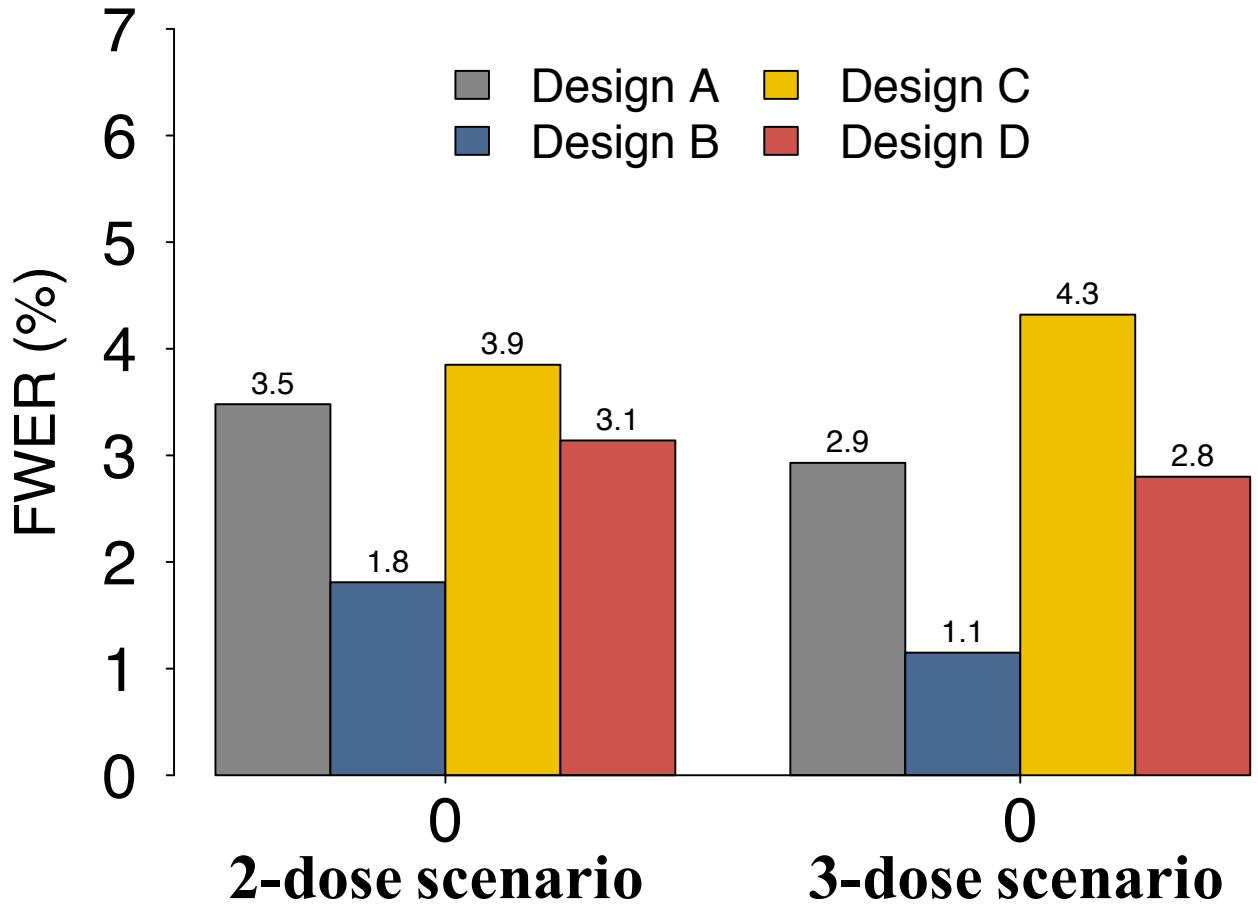

**Fig S2.** Familywise type I error rates (FWER, %) of all designs.

**D. Different null and target ORRs**

We considered different null and target ORRs 0.1 and 0.3, and a target hazard ratio 0.66. The toxicity upper limit $\phi_T = 0.3$ and the efficacy lower limit $\phi_E = 0.2$. The remaining simulation settings are the same as before. For Design A, the sample sizes in stage 1 and stage 2 are $n_1 = 50, n_2 = 70$ (for two doses) and $n_1 = 80, n_2 = 52$ (for three doses); for Design B, $n_1 = 50, n_2 = 90$ (for two doses) and $n_1 = 80, n_2 = 85$ (for three doses); for Design C, $n_1 = 50, n_2 = 35$ (for two doses) and $n_1 = 70, n_2 = 50$ (for three doses); and for Design D, $n_1 = 40, n_2 = 30$ (for two doses) and $n_1 = 70, n_2 = 20$ (for three doses). The simulation settings and results are shown in Figures S3 and S4.

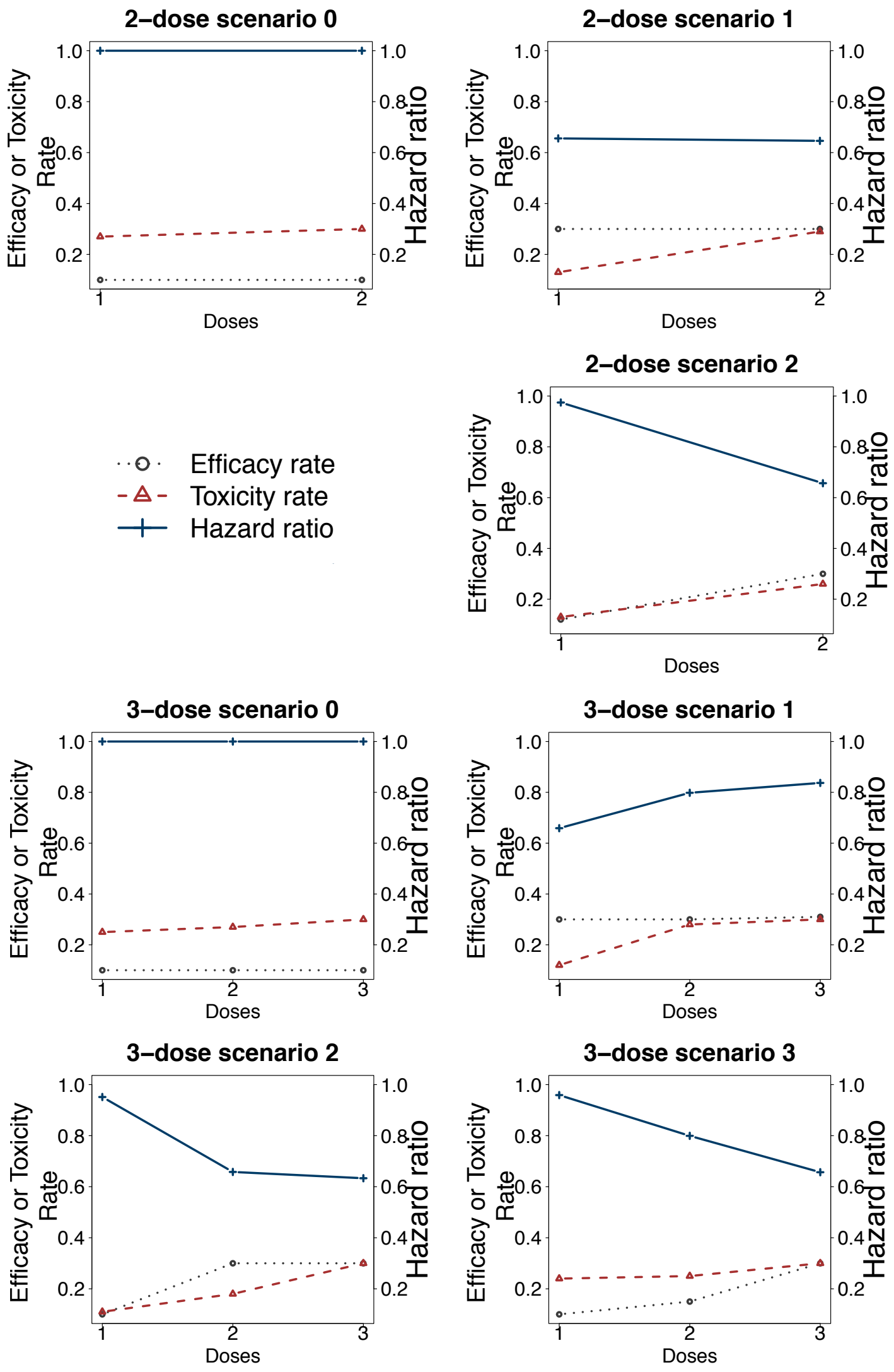

**Fig S3.** True response rates, toxicity rates, and hazard ratio for each dose in each scenario when two doses or three doses are evaluated in stage 1, with null and target efficacy rates of 0.1 and 0.3 and a target hazard ratio of 0.66.

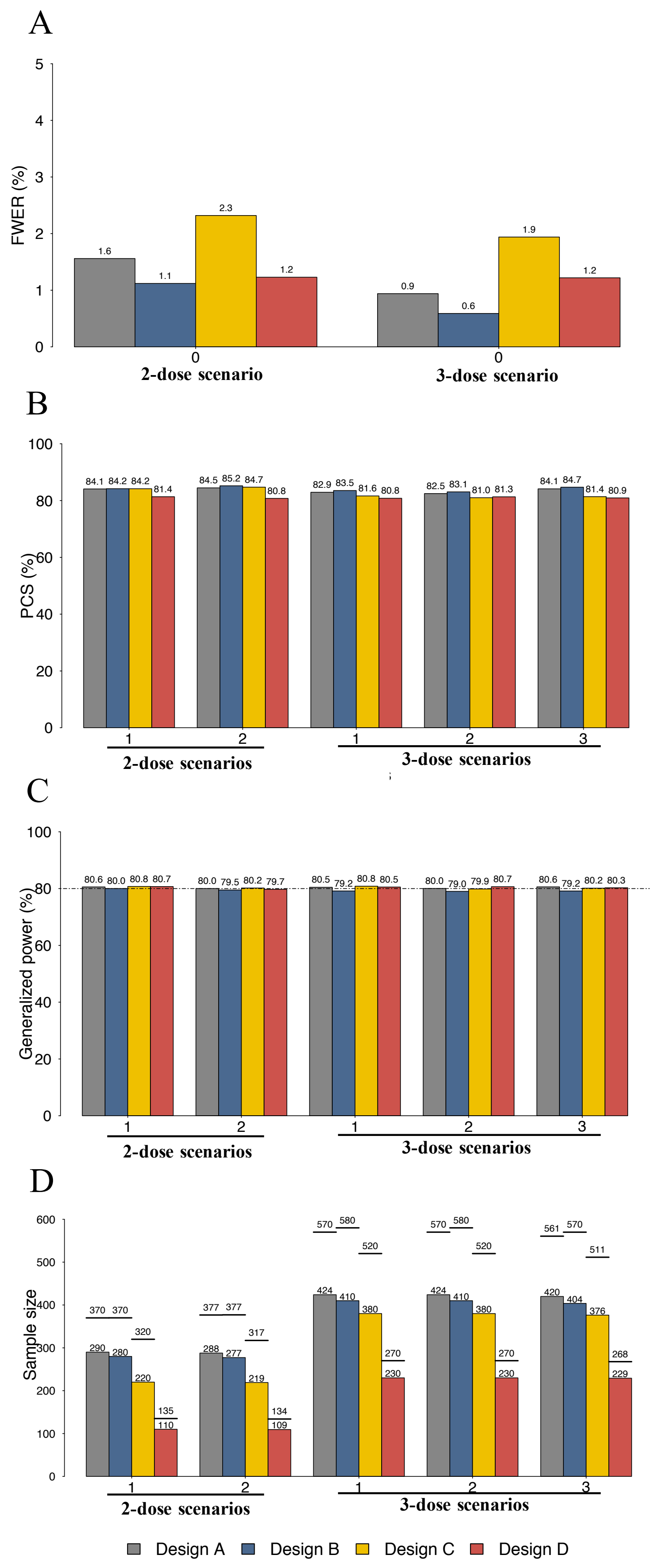

**Fig S4.** The familywise type I error rates (FWER, %), percentage of correct selection (PCS) of the optimal dose at stage 1, generalized power, and the average sample size of all designs and their conventional counterparts (CCs), with null and target ORRs of 0.1 and 0.3 and a target hazard ratio of 0.66. In panel (D), the horizonal lines indicate the sample size of conventional approach.

## E. Designs B and D with a patient population drift

Based on the previous simulation settings, we considered a population drift in two opposite directions. For Design B, in the case of a positive drift which leads to an overestimation of treatment effect, the efficacy rate and hazard of the historical control are 0.05 and 0.34. In the case of a negative drift which leads to an underestimation of treatment effect, the efficacy rate and hazard of the historical control are 0.35 and 0.22. For Design D, in the case of a positive drift, the efficacy rate of the historical control is 0.17. In the case of a negative drift, the efficacy rate of the historical control is 0.23. In the case of no population drift, the efficacy rate and hazard of the historical control are same as before, i.e., 0.2 and 0.26. The remaining simulation settings are same as before. The simulation results are shown in Figure S5.

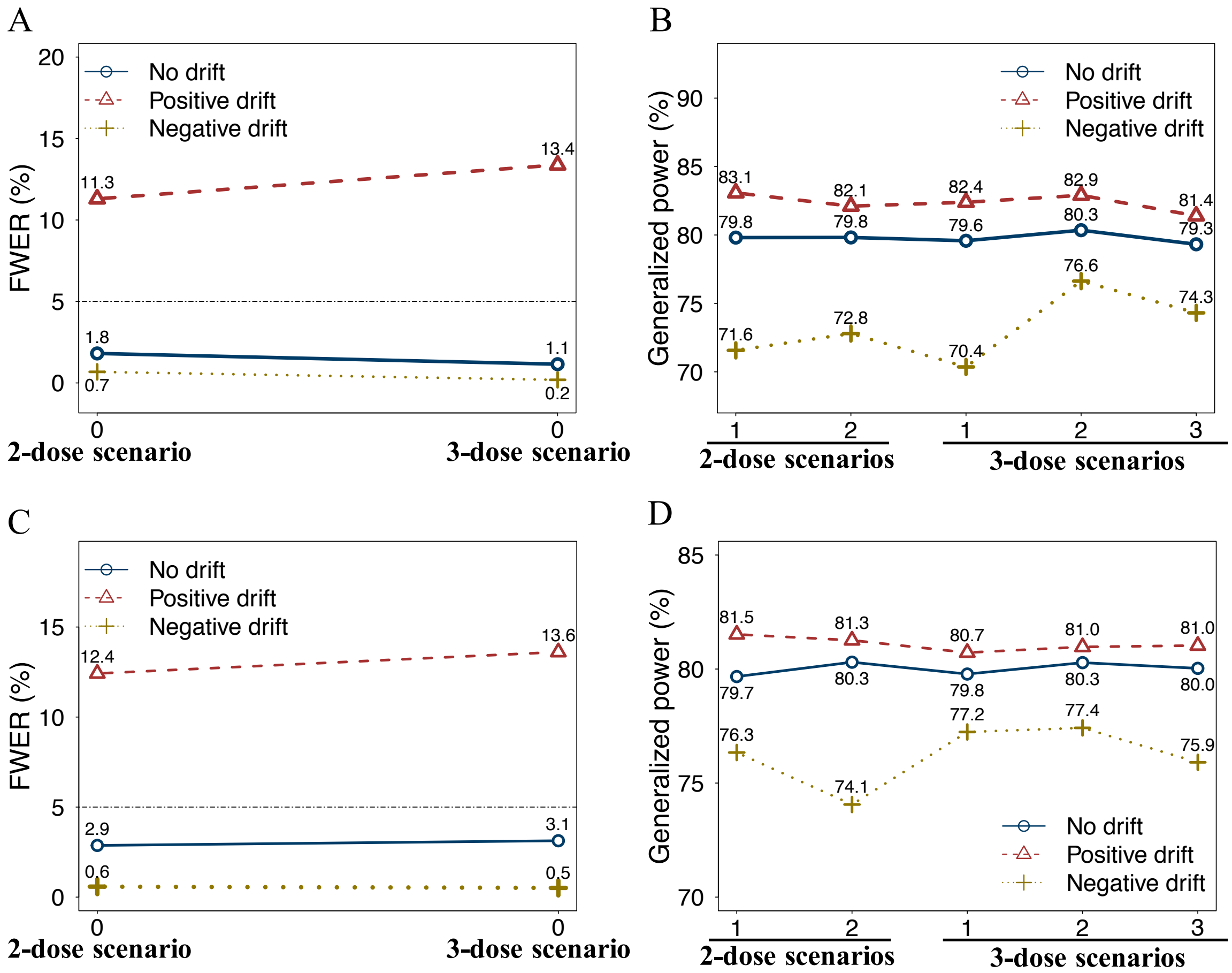

**Fig S5.** Familywise type I error rates (FWER) and generalized power of Design B (Figs S5A and S5B) and Design D (Figs S5C and S5D) in the case of no drift, positive drift, and negative drift.

## F. Study on different sample size at stage 1

The following Figure S6 shows the percentage of correct selection (PCS) of the optimal dose and generalized power of Design C, under different stage 1 sample size $n_1$ per arm in 2-dose scenario 2, 3- dose scenario 1 and 3-dose scenario 3.

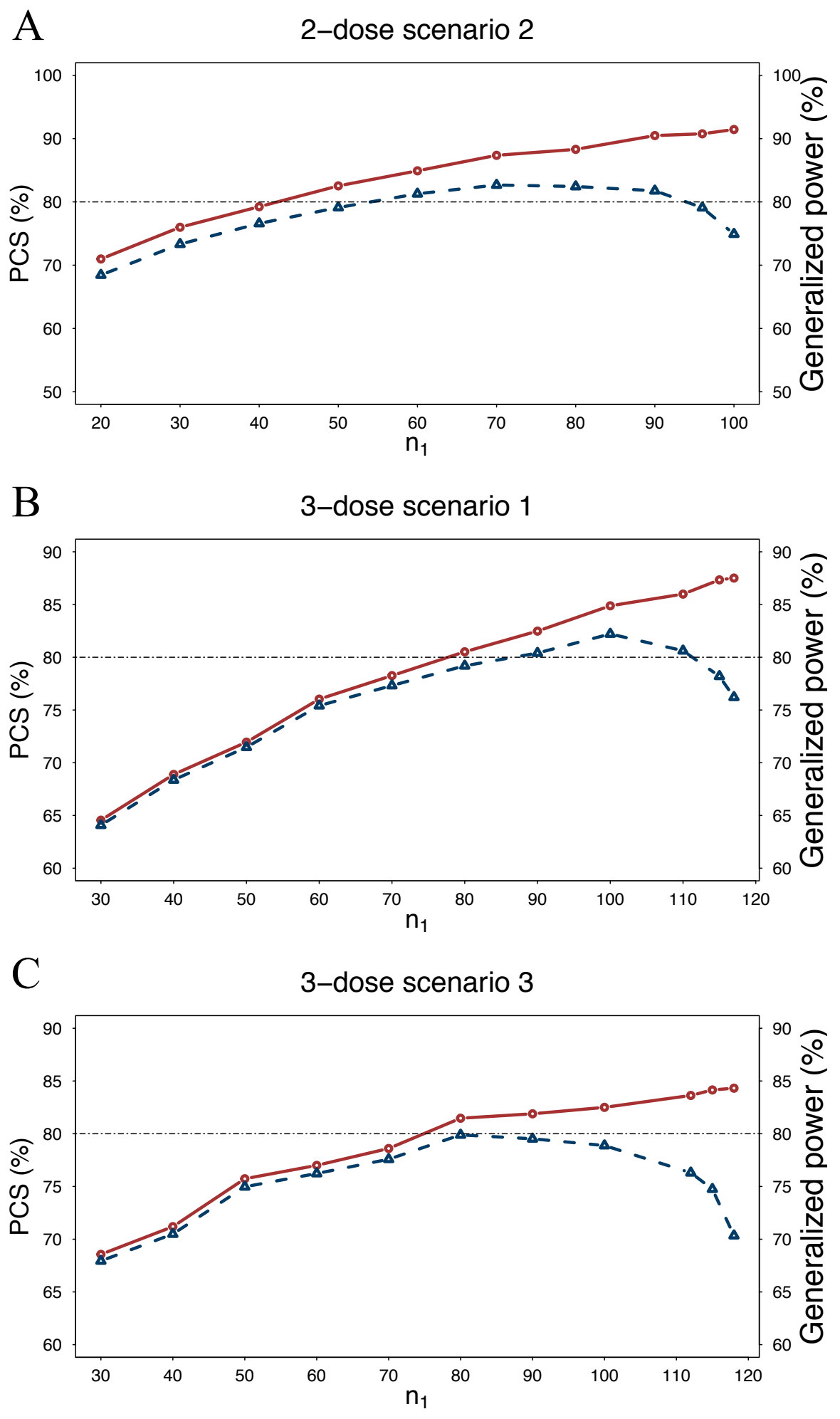

**Fig S6.** The percentage of correct selection (PCS, red solid lines) of the optimal dose and generalized power (blue dashed lines) of Design C, under different stage 1 sample size $n_1$ per arm in 2-dose scenario 2, 3-dose scenario 1 and 3-dose scenario 3. The total sample size is fixed.